# Shapley Value Methods for Attribution Modeling in Online Advertising

Kaifeng Zhao
Data and Analytics R&D, GroupM, Singapore
Kaifeng.zhao@groupm.com

Seyed Hanif Mahboobi
Data and Analytics R&D, GroupM, New York, NY, USA
Seyed.Mahboobi@groupm.com

Saeed R. Bagheri
Data and Analytics R&D, GroupM New York, NY, USA
Saeed.Bagheri@groupm.com

# Shapley Value Methods for Attribution Modeling in Online Advertising

**Abstract**

This paper re-examines the Shapley value method for attribution analysis in the area of online advertising. As a credit allocation solution in cooperative game theory, Shapley value method directly quantifies the contribution of online advertising inputs to the advertising key performance indicator (KPI) across multiple channels. We simplify the calculation method by developing an alternative mathematical formulation. The new formula significantly improves the computational efficiency and therefore extends the scope of applicability. Based on the simplified formula, we further develop the ordered Shapley value method. The proposed method is able to take into account the order of channels visited by users. We claim that it provides a more comprehensive insight by evaluating the attribution of channels at different stages of users' conversion journeys. The proposed approaches are illustrated using a real-world online advertising campaign dataset.

# 1  Introduction

In a digital age, advertisements can reach and influence users across various channels and devices. Online channels including search engines, publishers, digital video, etc., are popularly utilized in advertising campaign planning. They can deliver advertisement impressions to target users on PCs, smartphones, tablets, etc.  As a result, a user's decision to convert (purchase a product or sign up for a service) is usually driven by multiple touch points with advertisements. Therefore, a critical problem appealing to advertisers is how advertising key performance indicator (KPI) such as conversion or revenue should be attributed to all the advertising channels involved in a campaign. The systematic approaches to solve this problem are classified as attribution modeling (Kireyev et al., 2013; Kitts et al., 2010; Rentola, 2014; Sparkman and Locander, 1980).

Marketing practitioners start attribution analysis with simple rule-based models, such as last-touch (first-touch), linear, U-shaped and time decay models. They allocate credits according to pre-defined heuristic rules. However, these models have significant limitations (Dalessandro et al., 2012; Shao and Li, 2011). Conversely, the desired attribution model should be data-driven. The resulting credit allocation should be specific to the campaign, product/service being advertised and even the users involved in the campaign.

An alternative way to address the attribution problem is based on advanced data analytics methodologies, known as algorithmic approaches, which have attracted increasing attention over the past decade (Abhishek et al., 2012; Anderl et al., 2013; Berman, 2013; Dalessandro et al., 2012; Li and Kannan, 2013; Shao and Li, 2011; Tucker, 2012; Wiesel et al., 2011; Yadagiri et al., 2015; Zhang et al., 2014). Shao and Li (2011) proposed the "simple probabilistic model". This methodology examines, in addition to each individual channel's effect, the joint effect of all possible combinations of two, three or more channels. The authors suggest that in practice, the contributions from interactions beyond second order need not be considered. Subsequently, Dalessandro et al. (2012) developed a channel importance attribution method as a generalization of Shao and Li (2011). This method analyzes channel interactions up to full order, i.e. the joint effects of all advertising channels. The resulting approach is equivalent to the Shapley value method in cooperative game theory (Osborne and Rubinstein, 1994; Roth, 1988; Shapley, 1952; Winter, 2002).

Shapley value method is a general credit allocation approach in cooperative game theory. It is based on evaluating the marginal contribution of each player in a game. The credit assigned to each individual player, i.e. Shapley value, is the expected value of this marginal contribution over all possible permutations of the players. Shapley value method gained application in various fields (Ma et al., 2010; O'Brien et al., 2015; Song et al., 2016; Wei and Zhang, 2014). For example, Ma et al. (2010) adopted it to develop a profit-sharing mechanism for internet service provider (ISP) settlement in internet economics. In the context of advertising attribution, Shapley value treats an advertising campaign as a cooperative game, and advertising channels as players in the game. The channels work cooperatively to attract, influence and convert users. The Shapley value of each channel is calculated based on its contribution to the advertising KPI, which may include individual contributions and synergy with the rest.

In this paper, we re-examine the Shapley value method for attribution modeling in online advertising. Performance metrics utilized in this method are discussed and re-interpreted specifically in the advertising domain. Based on our interpretation, we develop a simplified calculation formula. The new formula is mathematically equivalent to the general Shapley value method but with significantly higher computational efficiency. We further propose an ordered Shapley value method to take into account the ordering effect of advertising channels. Our new approach can provide more insight into channels' advertising effectiveness at different stages of users' conversion journeys.

In Section 2 of this paper, we review the Shapley value method and discuss the advertising performance metrics, including contribution and marginal contribution measurements. We also describe the associated properties that ensure efficiency and fairness in credit allocation. Section 3 presents our simplified calculation formula, while the detailed mathematical proof is presented in the appendix. Next, in Section 4, we propose the ordered Shapley value method, including motivation and attribution mechanism. Section 5 demonstrates the performance of the proposed approaches using a real online advertising campaign dataset. We conclude our work with a brief discussion in Section 6.

## 2 Shapley Value Method

Shapley value method treats advertising channels as players in a cooperative game. It directly measures channels' contribution to the advertising KPI, and takes the mean of all added contributions as attribution values. Specifically, given a grand coalition $P$ consisting of advertising channels $\{x_1, x_2, \ldots, x_p\}$, we use a utility function $v(S)$ to describe the contribution of $S$ which denotes a coalition of channels. The Shapley value can be calculated using the following formula,

$$\phi_j = \sum_{S \subseteq P \setminus \{x_j\}} \frac{|S|!\,(p - |S| - 1)!}{p!} \left( v(S \cup \{x_j\}) - v(S) \right), j = 1, \ldots, p, \qquad (1)$$

where $|S|$ is the cardinality of coalition $S$ and the sum extends over all subsets $S$ of $P$ not containing channel $x_j$. $v(S \cup \{x_j\}) - v(S)$, which is also denoted as $M(j, S)$, is the marginal contribution of channel $x_j$ to the coalition $S$.

As noticed in Equation (1), Shapley value method takes the weighted average of its marginal contribution over all possible coalitions for each channel. The coalition's contribution and marginal contribution can be measured in multiple ways to gain insight from different aspects. It can be related to the number of purchases, the number of reaches as well as the total revenue (Berman, 2013). In the following sections, we discuss the definition of contribution and marginal contribution.

## 2.1 Contribution

The utility function $v(\cdot)$ measures the contribution of a channel coalition to the advertising KPI in the absence of the rest. For the grand coalition $P$ that includes all the channels in the campaign, $v(P)$ should equal to the total value created by the entire campaign. The campaign value is generated by all the converted users that have been reached by the advertisements. In reality, most converted users make decisions after visiting a particular subset of channels even though they are within reach of all available channels in $P$. If we define $u_j$ to be the set of users who have visited channel $x_j$ before conversion (they may or may not have visited other channels), the total campaign value is generated by the users is

$$U_P = \bigcup_{j \in P} u_j, \qquad (2)$$

which contains all the converted users. These users can be further grouped into multiple user types according to the channels they have visited. For example, user type $\{1\}$ contains all the users who have converted after visiting channel $x_1$ only, user type $\{2\}$ contains all the users who have converted after visiting channel $x_2$ only, user types $\{1,2\}$ contain all of the users who have converted after visiting both channel $x_1$ and $x_2$, and so on. Consequently, we can define the contribution of $P$ as

$$v(P) = \sum_{S \subseteq P} R(S), \qquad (3)$$

where $R(S)$ is the total value created by the users who have visited all the channels in $S$. We name it as the individual contribution of coalition $S$. Therefore, $v(P)$ contains the contributions from all of the converted users. It is also the total campaign value and the total credits to be allocated in our attribution problem.

Similarly, for each coalition $S \subseteq P$, the contribution of $S$ made by the users is

$$U_S = \bigcup_{j \in S} u_j \setminus \bigcup_{k \notin S} u_k. \qquad (4)$$

This means that the contribution of any coalition $S$ is measured as the value created by the users who have converted after visiting some channel(s) only in $S$. These users shall not visit any channel outside of $S$. Therefore, for any two coalitions $S_1$ and $S_2$ satisfying $S_1 \subseteq S_2$, we have,

$$U_{S_1} \subseteq U_{S_2}. \qquad (5)$$

Similar to ( 3 ), we define,

$$v(S) = \sum_{T \subseteq S} R(T). \tag{6}$$

In practice, $R(\cdot)$ could be the total number of reaches, total number of conversions or revenue/profit. We hereafter use revenue as our advertising KPI. Our analysis can be extended to other cases as well.

## 2.2 Marginal Contribution

We quantify the additional contribution of a channel ($x_j$) when added to a coalition ($S$) as its marginal contribution to $S$.

$$M(j, S) = v(S \cup \{x_j\}) - v(S). \tag{7}$$

In practice, this means we add a new channel to an existing campaign that contains all the channels in $S$. This new channel may likely bring new users who are not aware of the advertised products or services. It may also further influence the existing users who have been exposed to some of the channels in $S$ or convert the users directly regardless of their prior exposures. On the other hand, availability of additional channel does not necessarily mean that every member of the population will continue to be targeted by all existing channels plus this new channel. Some users will convert right after seeing this new channels before the other channels could reach them. A bulk of the users will not get an impression from this new channel because they convert before this new channel reaches them.

The contribution $v(S \cup \{x_j\})$ is made by the following three types of users.

- Type 1: Users who have visited some channel(s) in $S$ only.
- Type 2: Users who have visited channel $x_j$ only.
- Type 3: Users who have visited channel $x_j$ AND some channel(s) in $S$.

Type 1 users contribute to $v(S)$, which is part of $v(S \cup \{x_j\})$ and eventually part of $v(P)$. Type 2 users contribute to $M(j, S)$, i.e., the individual contribution from channel $x_j$. Lastly, type 3 users contribute to $M(j, S)$ as well, which is regarded as the advertising synergy among $x_j$ and the channels in $S$.

Note that the marginal contribution is generated by additional users (types 2 and 3). Therefore, $M(j, S)$ should always be non-negative.

## 2.3 Properties

A well-defined Shapley value approach should have certain desired properties (Ma et al. (2010)) such as symmetry, etc. It can be verified that our method defined as above satisfies all these properties.

- Efficiency

$$\sum_{j \in P} \phi_j = v(P). \tag{8}$$

This ensures that sum of Shapley values (attribution values) of channels equals to the total campaign value. Moreover, Shapley value method is a mechanism used to allocate total value to all players who have made a contribution. Therefore, we should ensure that every single credit allocated (Shapley value) comes from the total value.

- Dummy player: If channel $x_j$ is such that $M(j, S) = 0$ for every coalition $S$ not containing $x_j$, then

$$\phi_j = 0. \tag{9}$$

A channel that makes no contribution to any coalition will get zero credit. Since such channels cannot improve the cooperative performance, they are regarded as dummy players and can be removed from the game.

- Symmetry: If channels $x_{j_1}$ and $x_2$ are such that $M(j_1, S) = M(j_2, S)$ for any coalition $S \in P \setminus \{x_{j_1}, x_{j_2}\}$, then

$$\phi_{j_1} = \phi_{j_2}. \tag{10}$$

The symmetry property ensures that if two channels contribute the same to all possible coalitions, they will get the same Shapley values.

Lastly, we point out that our specification of Shapley value method is different from that in the literature (Dalessandro et al. 2012; Shao and Li, 2011). Specifically, Shao and Li (2011) proposed a simple probabilistic model. They defined the contribution of a coalition $S$ (a single channel or a pair of channels) using the value created by the users who have visited all channels in $S$. Dalessandro et al. (2012) further generalized this idea to a Shapley value approach. Consequently, the contribution of the grand coalition $P$ contains only the value created by the users who have visited all the channels in the campaign. However, our experience with real data suggests that users do not typically visit all available channels before conversion. According to the efficiency property, their definition may potentially lead to the analysis that focuses only on a small portion of the total campaign value. In addition, they interpreted that the marginal contribution of a channel, $M(j, S)$, is created by additional exposures to channel $x_j$ (after exposures from $S$). Therefore, the marginal contribution may be negative as the compared groups of users do not have the desired inclusion feature as in ( 5 ); i.e., an alternative group of users with exposures to an additional channel does not necessarily create higher value.

## 3 Simplified Shapley Value Method

The method we have described in the previous section is a direct implementation of the general Shapley value method. It suffers from heavy computation burden as we have to loop over all possible coalitions for each channel. For example, with 20 channels in a campaign, the total number of sub-coalitions is 1,048,576. Thus, implementing the original formula is not feasible for large-scale data analysis or evaluation of advertising performance in a timely manner. Therefore, we develop a simplified formula for Shapley value calculation. The new version can be regarded as a customized variation of the general Shapley value method for the attribution modeling in advertising.

Recall that we define multiple user types according to the channels they have visited. In other words, for each coalition $S$, we find all the users who have visited all the channels in $S$ before conversion. We use $R(S)$ to denote the total value created by these users.

As discussed in Section 2.2, the marginal contribution, $M(j, S)$, is made by the users who have visited channel $x_j$ plus a subset of $S$ (including the empty set). This motivates us to re-write the marginal contribution as following,

$$M(j, S) = \sum_{T: \{x_j\} \in T \, \wedge \, T \subseteq S \cup \{x_j\}} R(T), \quad j = 1, \dots, p. \quad (11)$$

By replacing the marginal contribution in ( 1 ) with ( 11 ), the Shapley value of any channel $x_j$ is simply a weighted average of $R(T)$ for all possible coalition $T$'s that contain channel $x_j$. The weight for each $T$ is calculated based on the following simplification:

**Theorem 1**: The Shapley values in a cooperative game defined in Section 2 can be calculated using the following simplified formula.

$$\phi_j = \sum_{S \subseteq P \setminus \{x_j\}} \frac{1}{|S| + 1} R(S \cup \{x_j\}), j = 1, \dots, p, \quad (12)$$

where $R(S \cup \{x_j\})$ denotes the revenue from the users who have visited all the channels in coalition $S \cup \{x_j\}$.

For example, for channel $x_1$ in a campaign with $P = \{x_1, x_2, x_3\}$, the Shapley value is,

$$\phi_1 = R(x_1) + \frac{1}{2} R(x_1, x_2) + \frac{1}{2} R(x_1, x_3) + \frac{1}{3} R(x_1, x_2, x_3). \quad (13)$$

The detailed mathematical proof is presented in the appendix.

A straightforward benefit of the new formula is that it greatly improves the computational efficiency. Instead of looping over all possible coalitions for each channel to calculate its marginal contribution, we evaluate each coalition at most once for each channel to directly calculate its Shapley value.

In addition to the higher computational efficiency, the new formulation helps to better understand the Shapley value method as following.

- The relationship between Shapley values and individual contribution of each coalition ($R(\cdot)$) is more evident: Shapley value of any channel is a weighted average of the individual contribution of all coalitions including the channel and takes no credit from any coalition excluding it.

- We can easily verify that Shapley values sum up to the total campaign outcome. The individual contribution of each sub-coalition is evenly distributed to all the channels included in this sub-coalition. Therefore, the sum of Shapley values equals to the sum of the individual contribution of all possible sub-coalitions, which is exactly the total campaign value (see ( 3 )).

- The individual contribution of any single channel $x_j$ is entirely allocated to $\phi_j$. This is because the weight of $R(x_j)$ is 1 in $\phi_j$ and is 0 in $\phi_k$ for any $k \neq j$.

- If a channel $x_j$'s marginal contribution to every coalition is 0, as $R(S \cup \{x_j\}) = 0$ for every $S$ not containing $x_j$. Therefore, $\phi_j$ is 0 as well.

- If any two channels $x_{j_1}$ and $x_{j_2}$ satisfy $M(j_1, S) = M(j_2, S)$ for every $S$ not containing $x_{j_1}$ and $x_{j_2}$, we have $R(S \cup \{x_{j_1}\}) = R(S \cup \{x_{j_2}\})$ for all such $S$'s. Therefore $\phi_{j_1} = \phi_{j_2}$.

## 4   Ordered Shapley Value Method

Based on the simplified formula in Section 3, we further extend the Shapley value method to incorporate the ordering effect of the channels visited by users. We call the resulting approach as ordered Shapley value method.

### 4.1   Motivation

A major problem with the general Shapley value method is ignoring the specific paths users may take to arrive at conversions. It treats channels indifferently regardless of their order of appearance in the conversion path. A channel is believed to have same attribution value as either the first or the last channel to reach the users.

However, marketing practitioners tend to believe that a channel may play different roles at different stages of a conversion process. At the early stages of a campaign, most users are not aware of the products or the services being advertised. The first several touchpoints will attract the users' initial attention. Subsequently, some users may start to follow the information about brand, products or services. Next, we activate the users' interests by further sending them advertisements across multiple devices and channels. Finally, some users make purchasing decisions at the conversion stage.

In practice, a channel may have different impacts on users' decision-making at different stages of the conversion journey. For example, a display channel naturally does not include very detailed information about the products or services. It can be a very effective introducer as the first touchpoint. However, it is less likely to directly convince the users to convert. On the contrary, the advertiser's official website can provide in-depth information to the users and can be effective in further convincing them, but may not be able to reach a large number of users at the early stages of a campaign. Therefore, it would be helpful if we could understand the role that every channel can play at each step of the conversion process and evaluate the corresponding attribution values.

## 4.2 Attribution Values

For each channel $x_j$ and a coalition $S \in P\setminus\{x_j\}$, users who contribute to $R(S \cup \{x_j\})$ may visit the channels in $S \cup \{x_j\}$ in different orders. We further differentiate them based on the location of channel $x_j$ in the sequence (touchpoint). Specifically, we use $R^i(S \cup \{x_j\})$ to denote the contribution made by the users who have visited all channels in $S \cup \{x_j\}$ as well as channel $x_j$ as the $i$th one during their conversion journeys. Here, $i$ is the index for different touchpoints in the sequence of channels. It can take values from $1, 2, \ldots, N$, where $N$ is the longest conversion journey for any user. Therefore,

$$R(S \cup \{x_j\}) = \sum_{i=1}^{|S|+1} R^i(S \cup \{x_j\}). \qquad (14)$$

It should be noted that, in case a user has visited a channel at multiple touchpoints, we evenly distribute the contribution to all the touchpoints.

Next, we define our ordered Shapley value method. Given a cooperative game as defined in Section 2, the ordered Shapley values can be calculated as follows,

$$\phi_j^i = \sum_{S \subseteq P\setminus\{x_j\}} \frac{1}{|S|+1} R^i(S \cup \{x_j\}), i = 1, \ldots, N, j = 1, \ldots, p, \qquad (15)$$

Furthermore, we define the following.

- Shapley value for channels

$$\phi_j = \sum_i \phi_j^i, j = 1, \ldots, p. \qquad (16)$$

It can be shown that this is equivalent to the original Shapley value method as discussed in Section 2.

- Shapley value for touchpoints

$$\phi^i = \sum_j \phi_j^i, i = 1, \ldots, N. \qquad (17)$$

This can provide new insights on how advertising efforts influence the users' decision-making at each step along their conversion journey.

## 5 Numerical Results

In this section, we use a real advertising campaign dataset to illustrate and demonstrate the performance of the proposed approaches.

### 5.1 Data Description

Our dataset is based on a 3-month online campaign which involves 18 advertising channels from 3 categories. Namely, they are Publishers, DSPs and Paid Search. The dataset records both impression and conversion history of each converted user. It includes a total of 153,814 revenue-generating observations. Table 1 presents a detailed description of the dataset.

Table 1: Data description.

|  | PUBLISHERS | DSP'S | PAID SEARCH | TOTAL |
| --- | --- | --- | --- | --- |
| Total Revenue ($) | 2182200.57 | 6150287.95 | 6759791.57 | 14313505.32 |
| Impressions | 101900 | 851333 | 97408 | 1050641 |

### 5.2 Group-wise Analysis

In this section, we evaluate the performance of the three advertising categories in the campaign. The aforementioned methods, including the original, simplified and ordered Shapley value methods, are utilized to calculate the attribution values. Note that the attribution values reported in the rest of this paper are percentages.

Figure 1 shows the attribution values obtained from Shapley value method for all categories. Since the two versions of formula produce similar numerical results, we report only one set of the attribution values. However, they differ substantially in computational efficiency: the analysis time is reduced from 17 hours to about 2 minutes on an ordinary PC using our simplified formula. As we can see in the chart, it is suggested that the paid search category has the highest attribution value at 47%. DSPs category is assigned a lower yet very close attribution value (40.43%). Lastly, publishers get the lowest value of 12.58%.

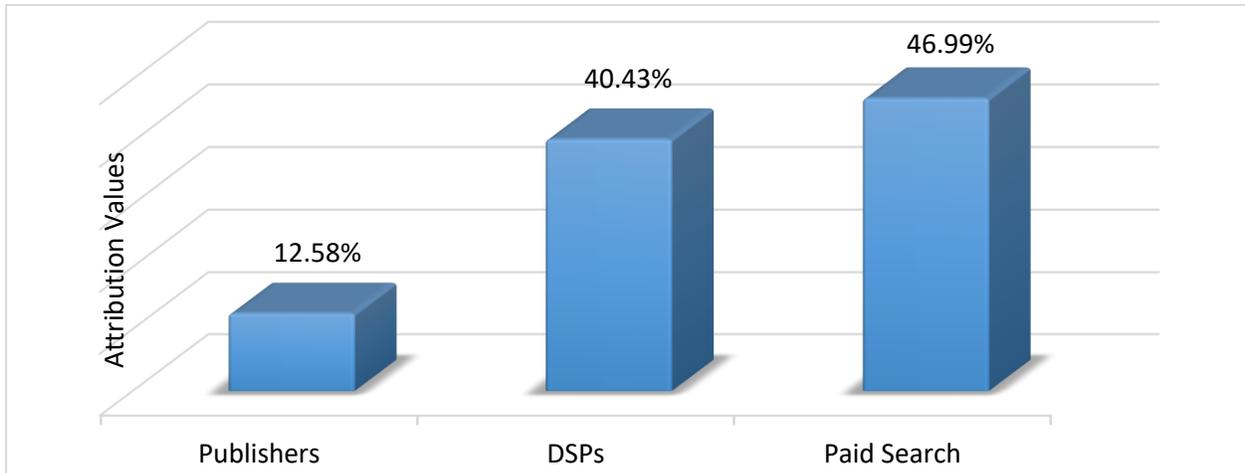

Figure 1: Attribution values for all channel categories.

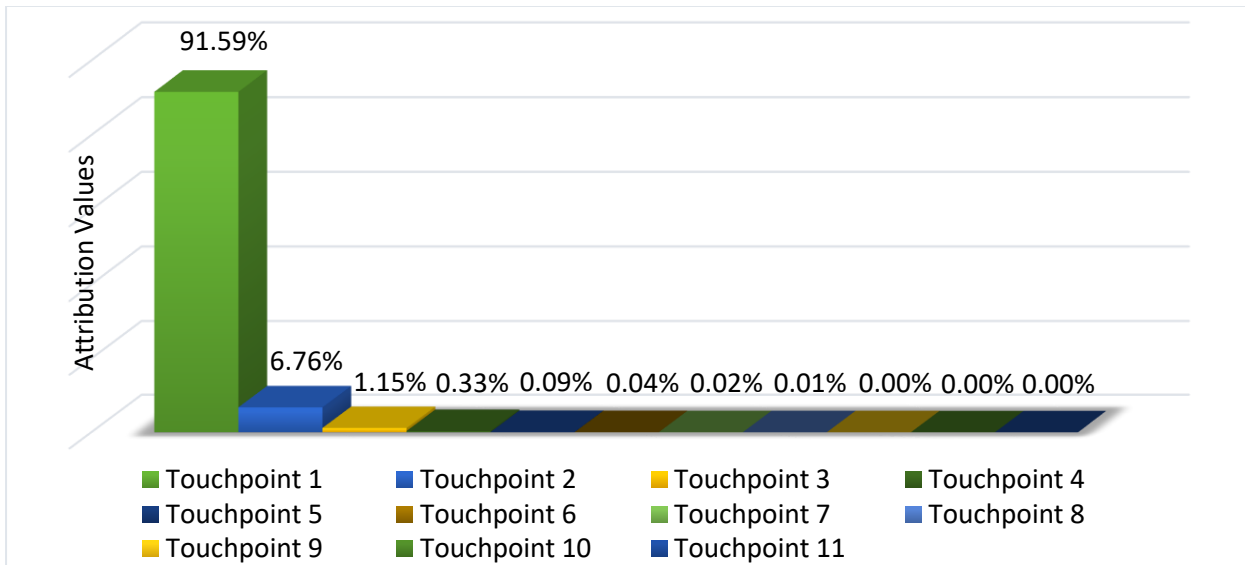

Figure 2: Attribution values for all touchpoints.

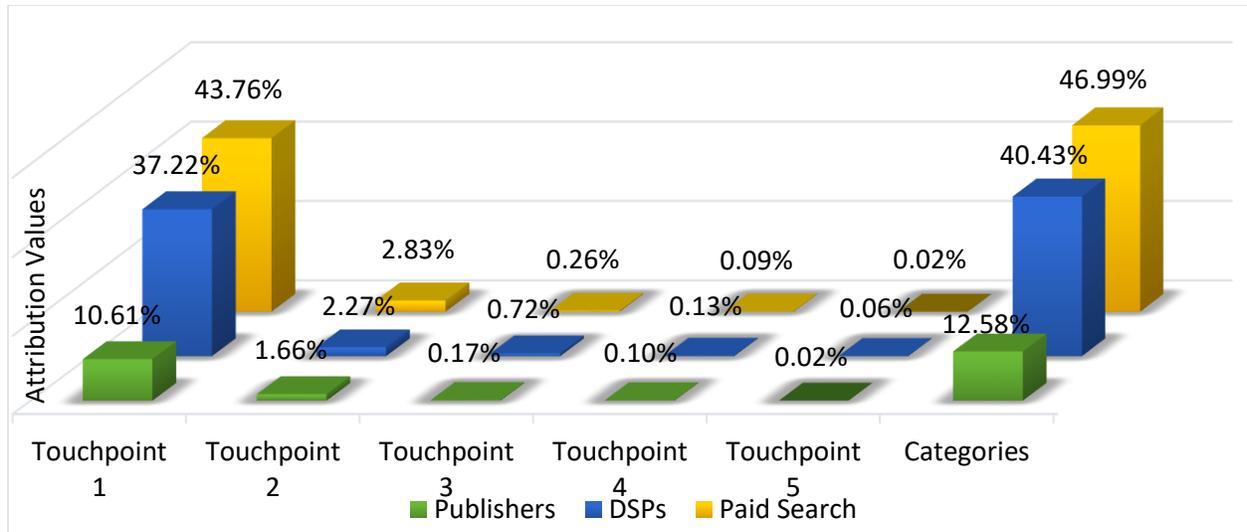

Figure 3: Attribution values for three channel categories at touchpoints 1-5.

Next, we apply our proposed ordered Shapley value method to investigate the attribution values of different touchpoints along users' conversion journeys. Specifically, attribution values for all touchpoints are calculated using Equation ( 17 ) and plotted in Figure 2. The longest conversion journey in the current dataset consists of 11 touchpoints. As shown in Figure 3, touchpoint 1 gets most of the credits (91.59%), and attribution value declines very fast as users visit more channels. In Figure 3, we further present attribution breakdown for channel categories at touchpoints 1-5, since the attribution value becomes negligible after touchpoint 5. For each category, touchpoint 1 always takes the most credits, and attribution values of the rest of the touchpoints are quite low. However, we notice that the ranking of categories varies over the stages of users' conversion journeys. At the early stage (touchpoints 1 and 2), paid search contributes the most, followed by DSPs and publishers. Then for touchpoints 3-5, DSPs play a more important role in further influencing the users who have had some initial exposures. We note that the total attribution values for channel categories are consistent with the general Shapley values in Figure 1. The numerical results are also reported in Table 2.

Table 2: Attribution values for three channel categories at all touchpoints.

| TOUCHPOINT | PUBLISHERS | DSP'S | PAID SEARCH | TOTAL |
|---|---|---|---|---|
| **TOUCHPOINT 1** | 10.606% | 37.217% | 43.763% | 91.59% |
| **TOUCHPOINT 2** | 1.658% | 2.270% | 2.835% | 6.76% |
| **TOUCHPOINT 3** | 0.166% | 0.718% | 0.265% | 1.15% |
| **TOUCHPOINT 4** | 0.105% | 0.133% | 0.093% | 0.33% |
| **TOUCHPOINT 5** | 0.020% | 0.058% | 0.016% | 0.09% |
| **TOUCHPOINT 6** | 0.016% | 0.017% | 0.007% | 0.04% |
| **TOUCHPOINT 7** | 0.006% | 0.010% | 0.004% | 0.02% |
| **TOUCHPOINT 8** | 0.004% | 0.005% | 0.001% | 0.01% |
| **TOUCHPOINT 9** | 0.001% | 0.002% | 0.001% | 0.00% |
| **TOUCHPOINT 10** | 0.001% | 0.001% | 0.000% | 0.00% |
| **TOUCHPOINT 11** | 0.000% | 0.000% | 0.000% | 0.00% |
| **TOTAL** | 12.583% | 40.432% | 46.985% | 100.00% |

Given the numerical results above, a question may arise here: as touchpoint 1 gets more than 90% of the total credit, does this imply that the first-touch model is a good fit? However, we find the contradictory with common sense to assign the highest attribution value to paid search at early stages of the campaign. To investigate this issue, we further analyze the attribution based on population segments. Specifically, we apply the ordered Shapley value method to two sub-groups of the users. The first group is obtained by filtering out those who convert after visiting only one channel. The resulting attribution values are plotted in Figure 4. An important observation here is that the attribution value of paid search at touchpoint 1 dramatically drops from 43.76% to 8.25%, while the attribution values of other touchpoints (especially touchpoint 2) increase considerably. This implies that the credit allocated to paid search at touchpoint 1 (as in Figure 3) mostly comes from the users who convert after visiting only a paid search channel. The second group includes the users whose conversion journeys contain more than two channels. In other words, we exclude the users who convert after visiting two channels. The results are shown in Figure 5, where we observe that attribution value of paid search at touchpoint 1 further drops to 4.37%, and the attribution values of touchpoints 3 to 5 as well as all the channels at these touchpoints continue to increase. Furthermore, in comparison to Figure 4, the attribution values of touchpoints 1 and 2 both decrease. Therefore, one can conclude that a certain portion of their credits is from the users who convert after visiting exactly two channels.

In practice, we may call users who need very few ad exposures to convert as loyal users. In current dataset, a considerably large number of users show loyalty. Some users even convert after visiting a single paid search channel. It is also likely that these users have been exposed to some offline channels prior to the online campaign. Thus, they start their online journeys with somewhat strong willingness to convert. As we do not have the relevant data at this point, most of the offline channels' credits are allocated to the first online channel that these users usually visit, which is mostly paid search in this dataset. Aside from the observed loyalty, we can see from Figure 5 that more than 75% of the total credits are allocated to publishers and DSPs, which play an important role at all touchpoints. On the contrary, paid search channels usually make the moderate contribution to the revenue, and get the attribution value of 24.43%.

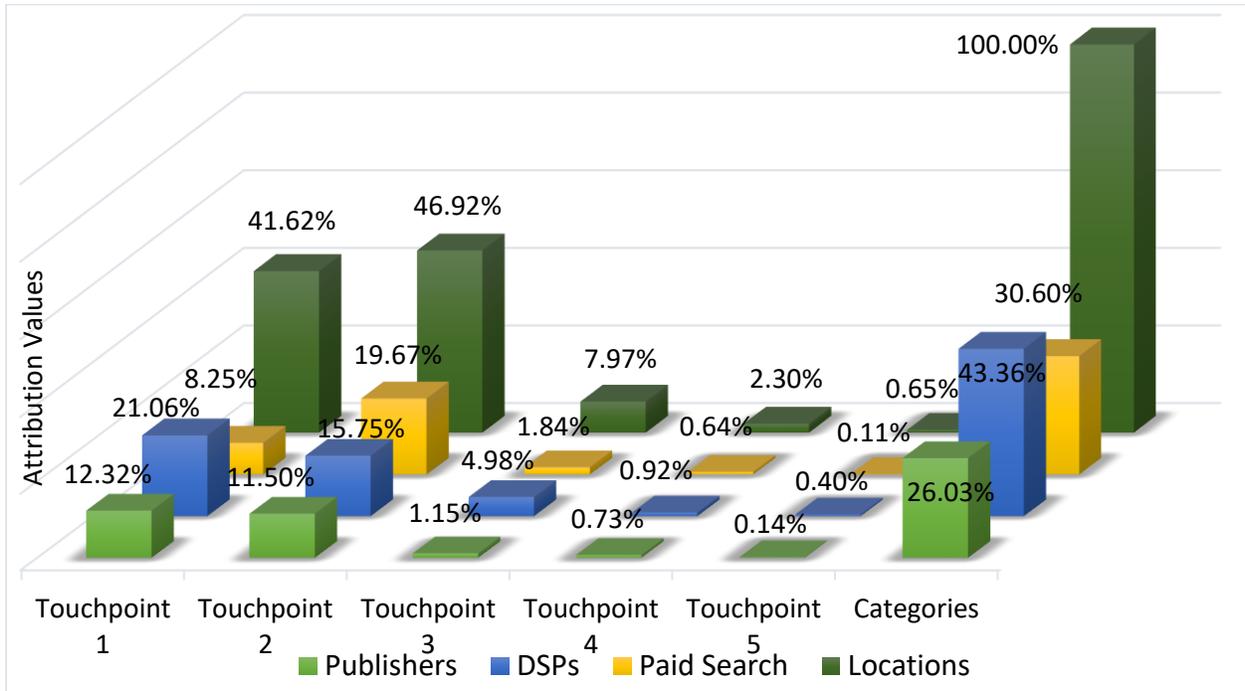

Figure 4: Attribution values based on users with conversion journey>1.

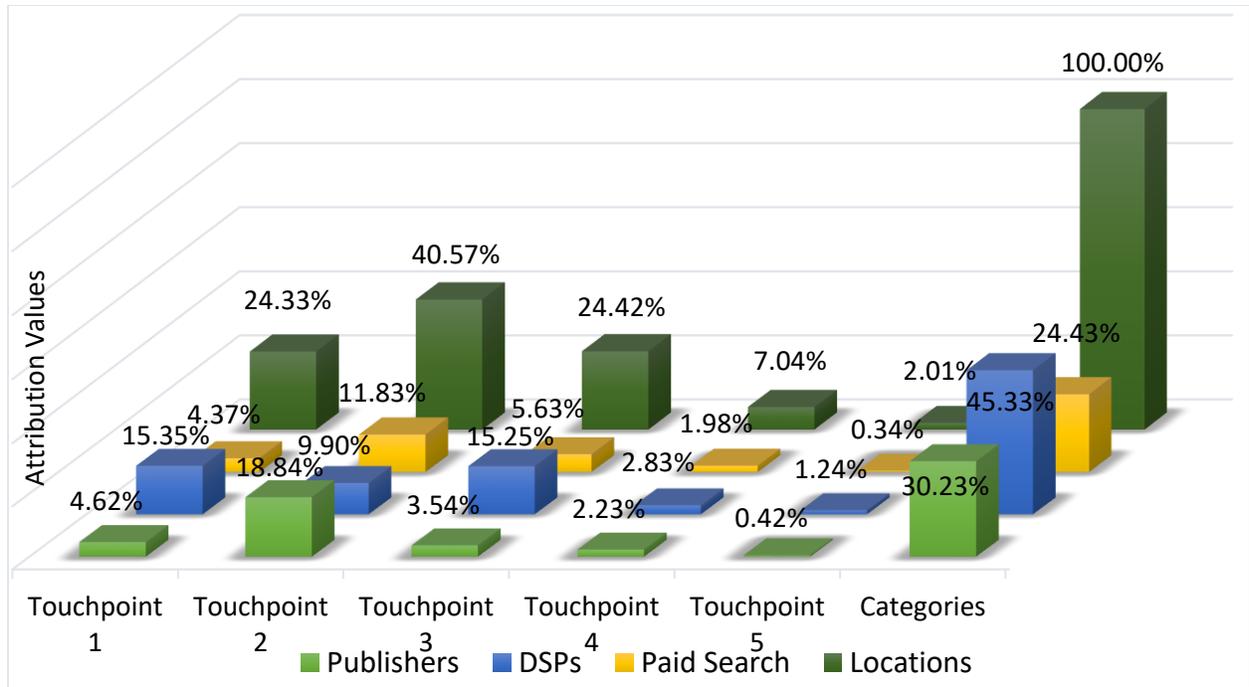

Figure 5: Attribution values based on users with conversion journey>2.

## 5.3 All-channel Analysis

In this section, we evaluate the attribution values for all 18 channels. As the proposed ordered Shapley value method is able to produce the same results as the general Shapley values (using Equation ( 16 )), we only focus on our new approach. The channels are named as P1, P2, etc. for those from publishers, D1 for a channel from a DSP, and S1 and S2 for channels from paid search.

In Table 3, we present touchpoint-specific attribution values for all the channels, as well as total attribution values for channels and touchpoints calculated using Equations ( 16 ) and ( 17 ) respectively. Similarly, we only report the results of the first 5 touchpoints. The analysis time takes about 2 minutes on an ordinary PC. The numerical results suggest that D1 is consistently the most important contributor at every touchpoint, followed by S1 and S2. On the contrary, the attribution values of most publishers are less than 1% which is negligible.

Table 4 shows the attribution values based on the users who visit more than two channels before conversion. As we can see from the table, D1 is again assigned the highest attribution value (still around 40%), while both paid search channels get fewer credits. This is due to the fact that loyal users mostly start their online journeys with search engines. Therefore, the credits of pre-campaign advertising efforts that help with building up the loyalty are mostly allocated to S1 and S2. After removing some loyal users, we observe that all publishers get higher attribution values. Channels including P9 and P12 are suggested to have close attribution values to S2. Also, instead of declining monotonically as in Table 3, the attribution values in Table 4 fluctuate along users' conversion journeys.

Table 3: Attribution values for all channels at touchpoints 1-5.

| CHANNEL | TOUCHPOINT 1 | TOUCHPOINT 2 | TOUCHPOINT 3 | TOUCHPOINT 4 | TOUCHPOINT 5 | TOTAL |
|---|---|---|---|---|---|---|
| P1 | 0.41% | 0.09% | 0.01% | 0.01% | 0.00% | 0.53% |
| P2 | 0.26% | 0.11% | 0.01% | 0.00% | 0.00% | 0.39% |
| P3 | 0.68% | 0.23% | 0.03% | 0.02% | 0.00% | 0.96% |
| P4 | 0.17% | 0.06% | 0.01% | 0.00% | 0.00% | 0.23% |
| P5 | 0.24% | 0.06% | 0.00% | 0.00% | 0.00% | 0.30% |
| P6 | 0.90% | 0.22% | 0.03% | 0.01% | 0.00% | 1.17% |
| P7 | 0.09% | 0.02% | 0.00% | 0.00% | 0.00% | 0.12% |
| P8 | 0.07% | 0.03% | 0.01% | 0.00% | 0.00% | 0.11% |
| P9 | 1.26% | 0.42% | 0.05% | 0.04% | 0.01% | 1.79% |
| P10 | 0.31% | 0.05% | 0.01% | 0.00% | 0.00% | 0.37% |
| P11 | 1.99% | 0.05% | 0.00% | 0.00% | 0.00% | 2.05% |
| P12 | 2.32% | 0.47% | 0.07% | 0.03% | 0.01% | 2.92% |
| P13 | 0.58% | 0.04% | 0.00% | 0.00% | 0.00% | 0.63% |
| P14 | 0.91% | 0.20% | 0.03% | 0.01% | 0.00% | 1.16% |
| P15 | 0.01% | 0.00% | 0.00% | 0.00% | 0.00% | 0.01% |
| D1 | 37.19% | 2.17% | 0.74% | 0.12% | 0.06% | 40.28% |
| S1 | 30.80% | 2.14% | 0.21% | 0.07% | 0.01% | 33.27% |
| S2 | 12.55% | 1.00% | 0.12% | 0.04% | 0.01% | 13.72% |
| TOTAL | 90.72% | 7.37% | 1.33% | 0.37% | 0.11% | 100.00% |

Table 4: Attribution values based on users with conversion journey>2.

| CHANNEL | TOUCHPOINT 1 | TOUCHPOINT 2 | TOUCHPOINT 3 | TOUCHPOINT 4 | TOUCHPOINT 5 | TOTAL |
|---|---|---|---|---|---|---|
| P1 | 0.17% | 0.84% | 0.21% | 0.15% | 0.03% | 1.43% |
| P2 | 0.02% | 1.33% | 0.21% | 0.09% | 0.01% | 1.73% |
| P3 | 0.28% | 2.10% | 0.49% | 0.31% | 0.04% | 3.30% |
| P4 | 0.03% | 0.36% | 0.10% | 0.04% | 0.02% | 0.58% |
| P5 | 0.11% | 0.55% | 0.08% | 0.05% | 0.01% | 0.82% |
| P6 | 0.30% | 2.17% | 0.48% | 0.26% | 0.06% | 3.34% |
| P7 | 0.08% | 0.24% | 0.05% | 0.03% | 0.01% | 0.41% |
| P8 | 0.01% | 0.21% | 0.10% | 0.05% | 0.01% | 0.40% |
| P9 | 0.58% | 3.82% | 1.00% | 0.68% | 0.11% | 6.39% |
| P10 | 0.06% | 0.46% | 0.14% | 0.06% | 0.01% | 0.75% |
| P11 | 2.78% | 0.70% | 0.06% | 0.01% | 0.00% | 3.56% |
| P12 | 1.57% | 4.93% | 1.40% | 0.62% | 0.21% | 8.86% |
| P13 | 0.09% | 0.14% | 0.08% | 0.02% | 0.00% | 0.34% |
| P14 | 0.15% | 1.48% | 0.61% | 0.26% | 0.07% | 2.63% |
| P15 | 0.00% | 0.00% | 0.00% | 0.00% | 0.00% | 0.01% |
| D1 | 14.05% | 7.59% | 13.94% | 2.34% | 1.10% | 39.63% |
| S1 | 3.15% | 8.51% | 3.91% | 1.30% | 0.26% | 17.29% |
| S2 | 1.66% | 3.54% | 2.30% | 0.79% | 0.13% | 8.53% |
| TOTAL | 25.07% | 38.97% | 25.17% | 7.06% | 2.09% | 100.00% |

# 6   Conclusion and Discussion

In this paper, we implemented an alternative formulation of Shapley value method. Performance metrics of campaign and channel coalition are proposed and discussed. The methodology ensures properties like efficiency and fairness. The resulting approach allocates the total campaign value to all the involved channels. Our simplified calculation formula leads to a customized variation for attribution analysis in the advertising domain. The resulting high computational efficiency can greatly extend its application scope to big data and omnichannel analysis. We also proposed ordered Shapley value method. This method is able to evaluate the attribution of advertising inputs at different stages of users' conversion journey. The new insights that the ordered methodology provides helps to draw a complete picture for advertising KPI attribution. Moreover, the ordered Shapley value method can be readily generalized to a time-dependent attribution approach with a slightly higher computational burden. This is achieved by attributing advertising KPI to the actual time stamps (hours, days, etc.) instead of touchpoints.

We note that the attribution analysis carried out in the paper has certain limitations. It is suspected that the abnormally high attribution value assigned to paid search at the early stage of the campaign partially comes from the offline channels. In other words, lack of offline data makes it difficult to correctly quantify the contribution of online efforts. Therefore, there is a need for a data integration solution that integrates both online and offline channels, which is beyond the scope of the current study. Nevertheless, we partially addressed this limitation by filtering out some 'loyal' users.

An area of future development is to incorporate advertising adstock in the ordered Shapley value method. This enables us to take into account such carry-over effect when evaluating the attribution value for touchpoints. Another point of improvement is to connect the Shapley value methods with advertising campaign planning or cross-channel optimization. As discussed in Ma et al. (2010), a Shapley value mechanism naturally incents players in a cooperative game to adopt a globally optimal strategy. The players' selfish behavior eventually leads to maximizing the value of the entire game. In the context of attribution in advertising, this means that an advertising plan that maximizes the total campaign value automatically maximizes the Shapley value of (credits assigned to) each channel, i.e., a Nash equilibrium solution. This can be easily verified using our findings in this paper (Equations ( 3 ) and ( 12 )). However, we note that this conclusion holds only when there is no budget restriction. It will be beneficial to combine attribution analysis with predictive modeling, and develop an effective budget allocation strategy in the proposed game theory framework.

# Appendix

In this section, we first present and prove a lemma, followed by the proof of Theorem 1.

**Lemma**: $n$ is a positive integer. For any integer $n_0 \in [1, n]$, we have,

$$\sum_{N=n_0}^{n-1} \frac{N!\,(n-n_0-1)!}{n!\,(N-n_0)!} = \frac{1}{n_0+1} \qquad (18)$$

*Proof*:

$$\sum_{N=n_0}^{n-1} \frac{N!\,(n-n_0-1)!}{n!\,(N-n_0)!}$$

$$= \sum_{N=n_0}^{n-1} \left[\frac{N!}{(N-n_0)!\,(n_0+1)!} \cdot \frac{(n-n_0-1)!\,(n_0+1)!}{n!}\right]$$

$$= \sum_{N=n_0}^{n-1} \frac{1}{n_0+1} \cdot \frac{N!}{(N-n_0)!\,n_0!} \cdot \frac{1}{C_n^{n_0+1}} = \frac{1}{n_0+1} \cdot \frac{\sum_{N=n_0}^{n-1} C_N^{n_0}}{C_n^{n_0+1}}$$

$$= \frac{1}{(n_0+1)C_n^{n_0+1}}\left(C_{n_0}^{n_0} + C_{n_0+1}^{n_0} + C_{n_0+2}^{n_0} + \cdots + C_{n-1}^{n_0}\right)$$

$$= \frac{1}{(n_0+1)C_n^{n_0+1}}\left(\frac{1}{n_0+1}C_{n_0+1}^{n_0} + C_{n_0+1}^{n_0} + C_{n_0+2}^{n_0} + \cdots + C_{n-1}^{n_0}\right)$$

$$= \frac{1}{(n_0+1)C_n^{n_0+1}}\left(\frac{n_0+2}{n_0+1}C_{n_0+1}^{n_0} + C_{n_0+2}^{n_0} + \cdots + C_{n-1}^{n_0}\right)$$

$$= \frac{1}{(n_0+1)C_n^{n_0+1}}\left(\frac{n_0+2}{n_0+1}\cdot\frac{2}{n_0+2}C_{n_0+2}^{n_0} + C_{n_0+2}^{n_0} + \cdots + C_{n-1}^{n_0}\right)$$

$$= \frac{1}{(n_0+1)C_n^{n_0+1}}\left(\frac{n_0+3}{n_0+1}C_{n_0+2}^{n_0} + \cdots + C_{n-1}^{n_0}\right)$$

… …

$$= \frac{1}{(n_0+1)C_n^{n_0+1}}\cdot\frac{n}{n_0+1}C_{n-1}^{n_0}$$

$$= \frac{1}{(n_0+1)}\cdot\frac{(n-n_0-1)!\,(n_0+1)!}{n!}\cdot\frac{n}{n_0+1}\cdot\frac{(n-1)!}{n_0!\,(n-n_0-1)!}$$

$$= \frac{1}{(n_0+1)}$$

**Proof of Theorem 1**:

For any channel $x_j$, $\phi_j$ is a linear combination of $R(S \cup \{x_j\})$'s with $S \subseteq P \backslash \{x_j\}$. This is because, as discussed in Section 3, every marginal contribution, $M(j, S)$, is a sum of $R(S' \cup \{x_j\})$'s with $S' \subseteq S$.

Then for any particular coalition $S_0 \subseteq P \backslash \{x_j\}$ with $|S_0| = p_0$, we want to evaluate the weight of $R(S_0 \cup \{x_j\})$ in $\phi_j$. As $R(S_0 \cup \{x_j\})$ belongs to all possible $M(j, S)$'s with $S_0 \subseteq S \subseteq P \backslash \{x_j\}$, its weight is simply the sum of the weights of such $M(j, S)$'s in the original formula.

Therefore, its weight equals to,

$$\sum_{S: S_0 \subseteq S \subseteq P \backslash \{x_j\}} \frac{|S|!\,(p - |S| - 1)!}{p!}$$

$$= \sum_{p_S = p_0}^{p-1} \frac{p_S!\,(p - p_S - 1)!}{p!} \cdot C_{p - p_0 - 1}^{p_S - p_0} = \sum_{p_S = p_0}^{p-1} \frac{p_S!\,(p - p_0 - 1)!}{p!\,(p_S - p_0)!},$$

where $p_S$ denotes the cardinality of $S$, i.e., $|S|$. $C_{p - p_0 - 1}^{p_S - p_0}$ is the total number of $S$'s that satisfy $S_0 \subseteq S \subseteq P \backslash \{x_j\}$. This is calculated by selecting additional $|S| - |S_0| = p_S - p_0$ channels from a pool of $p - |S_0| - |\{x_j\}| = p - p_0 - 1$ available channels, and adding them to $S_0$ to form a valid $S$. The sum extends over all possible cardinalities of $S$.

According to the Lemma, we can further simplify the above formula as,

$$\sum_{S: S_0 \subseteq S \subseteq P \backslash \{x_j\}} \frac{|S|!\,(p - |S| - 1)!}{p!} = \sum_{p_S = p_0}^{p-1} \frac{p_S!\,(p - p_0 - 1)!}{p!\,(p_S - p_0)!} = \frac{1}{p_0 + 1}$$

In other words, the weight of any $S_0$ in $\phi_j$ is simply $\frac{1}{|S_0|+1}$. As this holds for any arbitrary $S_0 \subseteq P \backslash \{x_j\}$, we have derived the final formula as

$$\phi_j = \sum_{S \subseteq P \backslash \{x_j\}} \frac{1}{|S| + 1} R(S \cup \{x_j\}) \tag{19}$$